\documentclass[prl,letterpaper,twocolumn,showpacs,superscriptaddress,floatfix]{revtex4}

\usepackage{graphicx,psfrag,amsmath,amssymb,amsfonts,bbm,latexsym,color,dcolumn,bm}

\begin{document}

\title{Reply to Comment on ``Anomalies in electrostatic calibrations
  for \\ the measurement of the Casimir force in a sphere-plane geometry''}

\author{W.J. Kim$^*$}
\affiliation{Department of Physics and Astronomy,Dartmouth College,6127 Wilder Laboratory,Hanover,NH 03755,USA}

\author{M. Brown-Hayes}
\affiliation{Department of Physics and Astronomy,Dartmouth College,6127 Wilder Laboratory,Hanover,NH 03755,USA}

\author{D.A.R. Dalvit}
\affiliation{Theoretical Division,MS B213,Los Alamos National Laboratory,Los Alamos,NM 87545,USA}

\author{J.H. Brownell}
\affiliation{Department of Physics and Astronomy,Dartmouth College,6127 Wilder Laboratory,Hanover,NH 03755,USA}

\author{R. Onofrio}
\affiliation{Dipartimento di Fisica ``Galileo Galilei'',Universit\`a  di Padova,Via Marzolo 8,Padova 35131,Italy}

\affiliation{Department of Physics and Astronomy,Dartmouth College,6127 Wilder Laboratory,Hanover,NH 03755,USA}

\begin{abstract}
In a recent Comment, Decca {\it et al.} have discussed the origin of
the anomalies recently reported by us in [Phys. Rev. A \textbf{78},
036102(R) (2008)]. Here we restate our view, corroborated by their 
considerations, that quantitative geometrical and electrostatic 
characterizations of the conducting surfaces (a topic not discussed 
explicitly in the literature until very recently) are critical for 
the assessment of precision and accuracy of the demonstration of the 
Casimir force, and for deriving meaningful limits on the existence 
of Yukawian components possibly 
superimposed to the Newtonian gravitational interaction.  
\end{abstract}


\maketitle
In the last decade, various efforts have been focused on demonstrating
the Casimir force and exploring hypothetical short-range forces of 
gravitational origin \cite{Fischbach}. Limits to the existence of 
these forces - or their tentative discovery - in the micrometer 
range rely on the control at the highest level of accuracy of the 
Casimir force and the related systematic effects \cite{Reynaudrev,Onofrio}. 

In this context we have investigated the celebrated sphere-plane
geometry in a range of parameters for which the hypothetical Yukawian 
contribution of gravitational origin should be optimally detected \cite{prarckim}. 
This implies exploiting a combination of spheres with large radius 
of curvature, such as the one already used in \cite{Lamoreaux}, and 
small separation gaps between the sphere and the planar surface, 
similar to the ones explored in \cite{Mohideen,Chan,Decca} with 
spheres having order of 100 $\mu$m radius of curvature. 
Notice that large radius of curvature and relatively large 
distances as in \cite{Lamoreaux} are not adequately sensitive 
to Yukawian forces with small interaction range. 
Conversely, microspheres at small distances as used in 
\cite{Mohideen,Chan,Decca} have small sensitivity to the 
amplitude of Yukawa forces, due to the smaller expected signal arising 
from the reduced effective surfaces of interaction. In this regard, 
limits to the Yukawa force based on a formal mapping between an
ideal parallel plate geometry and the sphere-plane configuration
actually used in the experimental setup as in \cite{DeccaAnn} are
invalid, as the Proximity Force Approximation (PFA), typically 
used for forces acting between surfaces \cite{Blocki}, does 
not hold for forces of volumetric character such as the gravitational 
force or its hypothetical short-range relatives. 

In \cite{prarckim}, we reported two anomalies in the electrostatic 
calibration of our apparatus, after discussing and ruling out some systematic effects. 
This has triggered the interest of the authors of \cite{DeccaComment} 
who have added two interesting points, first attempting to explain  
our first anomaly in terms of a systematic deviation from the ideal, 
single-curvature pattern for the spherical surface, and second presenting 
a distance independent contact potential in one of their experimental setups. 
We welcome these different insights and would like to discuss here
their implications in the general context of both accurate demonstrations 
of the Casimir force and precision experiments on Yukawian
gravitational forces, as in the following.

\textit{Deviation from ideal spherical geometry}: 
Among the possible reasons for the first anomaly we have briefly 
discussed in \cite{prarckim}, a couple of possibilities arise from 
geometrical effects, namely the validity of the PFA approximation 
in our case and a surface obtained by convolving various 
spheres with different radii of curvature having in common the 
point of minimum distance from the plate.
The authors of \cite{DeccaComment} provide a further example of 
a deviation from geometry that certainly, for an appropriate choice 
of the additional parameters added to the model, may mimic any 
desired power law exponent for the scaling of the electrostatic 
curvature coefficient $k_{\rm el}$ upon the distance. 
In particular, we concur that a softer dependence is obtainable 
by guessing higher curvatures around the point of minimum
distance. In general, it is indeed intuitive that for instance harder 
scaling with the distance is obtained in regions at lower curvature 
- in the extreme case of a flat surface ({\it i.e.} infinite radius 
of curvature) one expects a scaling of $k_{\rm el}$ with distance 
through an exponent -3 - and conversely exponents softer than 
the expected -2 should be associated with even {\it sharper} regions. 

However, this interpretation in terms of a modified geometry 
is hard to reconcile with the measurement of the capacitance 
versus distance that better follows the behaviour expected 
for a surface with a single radius of curvature, as shown in 
Fig. 1. The electrostatic curvature coefficient $k_\mathrm{el}$ 
is related to the second spatial derivative of the capacitance $C$  
and the effective mass $m_\mathrm{eff}$ as
\begin{equation}
k_\mathrm{el}=\frac{C''}{8 \pi^2 m_\mathrm{eff}}.
\end{equation}
The model considered in \cite{DeccaComment} implies a capacitance for 
the modified sphere-plane configuration expressed by the formula:
\begin{eqnarray}
C_\mathrm{mod} &=& 2 \pi \epsilon_0 [R_{CD}\ln(R_{CD}/d) 
\nonumber \\ 
&+&(R_{AB}-R_{CD})\ln\left(\frac{R_{AB}-R_{CD}}{d+h}\right)
\nonumber \\
&-&(R_{AB}-R)\ln\left(\frac{R_{AB}-R}{d+h+H}\right)]
\end{eqnarray}
(see \cite{DeccaComment} for the definition of the various geometrical
quantities) up to a term of the type $A_1+A_2 d$ arising from the double integration
of Eq. 1. This expression may be approximated, in the distance 
range discussed in \cite{DeccaComment}, as
\begin{equation}
\tilde{C}_\mathrm{mod}=A_1^\mathrm{mod}+A_2^\mathrm{mod} d+ A_3^\mathrm{mod}d^{0.3}.
\end{equation}
For the ideal sphere-plane capacitance we obtain instead 
\begin{equation}
C_\mathrm{id}=A_1^\mathrm{id}+A_2^\mathrm{id} d+ A_3^\mathrm{id} \ln(R/d),
\end{equation}
with $A_3^\mathrm{id}=-2\pi \epsilon_0 R$. 
The linear terms $A_2^\mathrm{id}$ and $A_2^\mathrm{mod}$ may represent the 
effect of a constant electric field, to which our AC capacitance meter 
should not be sensitive (see comment below). 
We have fitted our capacitance data with both Eqs. (3) and (4),  
supposing that $A_2^\mathrm{id}=A_2^\mathrm{mod}=0$. Fitting 
with the modified geometry (Eq. 3) yields a best fit with a reduced $\chi^2$
that is significantly larger than that of the ideal geometry (Eq. 4), as detailed in the
caption of Fig. 1. Moreover, by using the parameters 
provided in \cite{DeccaComment} ($R_{AB}$=1.6 R=49.4 mm, $R_{CD}$=30 
$\mu$m, $H$=250 nm, $h$=8 nm), chosen to reproduce the anomalous scaling 
power law observed by us in \cite{prarckim}, we find that the best  
fit with Eq. 2 gives a discrepancy of about 20 $\%$ in the expected coefficient 
for the distance dependence, while the discrepancy is 2.1 $\%$ using
the ideal sphere-plane formula, quite close to the nominal relative error of the 
radius of curvature of the sphere, $R=(30.9 \pm 0.15)$ mm. 
This suggests that the capacitance data are better explained 
by using an idealized geometry, rather than the sophisticated 
geometry proposed in \cite{DeccaComment} that, moreover, should have been 
evidenced by the dedicated AFM imaging of the sphere that we performed after
the runs. Notice that in our setup the capacitance is measured 
{\it dynamically} by using an AC bridge, while $k_\mathrm{el}$ is 
measured by looking at the frequency shift related to the gradient 
of  spatial-dependent but {\it static} force. By supposing that static 
or slowly-changing charges - not affecting the dynamical measurement 
of the capacitance - are present on the two surfaces, an ideal 
capacitance and an anomalous scaling of $k_\mathrm{el}$ are mutually consistent.

\begin{figure}[t]
\begin{center}
\includegraphics[width=0.90\columnwidth]{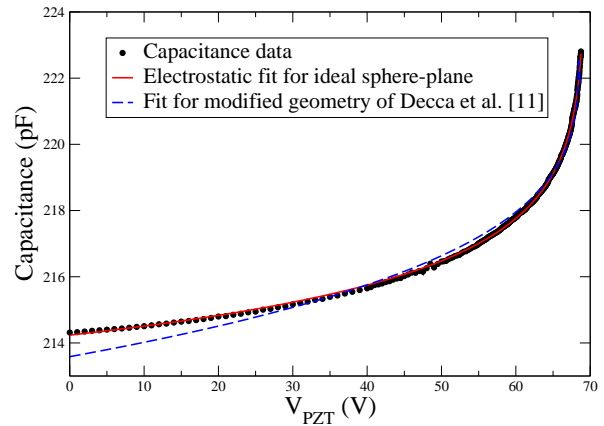}
\end{center}
\caption{(Color on-line) Capacitance versus PZT voltage data (black circles) 
and its best fit under the two hypotheses of ideal spherical geometry
and modified geometry as in \cite{DeccaComment}. 
The modified geometry fit (blue dashed line) is based on Eq. 3, with 
the distance $d=\beta(V^0_{\rm PZT}-V_{\rm PZT})$, where 
$\beta=(87 \pm 2)$ nm/V is the PZT actuation coefficient. 
The best fit with the constraint of $A_2$-terms equal to zero yields  
$A_1^\mathrm{mod}=(222.96 \pm 0.04)$ pF, 
$A_3^\mathrm{mod}=-(346.2 \pm 1)$ pF/m, 
$V^0_\mathrm{PZT}=(68.43 \pm 0.05)$ V, with a reduced $\chi^2=77.4$.
The ideal geometry fit (red continuous line) is based on Eq. 4, 
and the resulting parameters are
$A_1^\mathrm{id}=(193.9 \pm 0.2)$ pF, 
$A_3^\mathrm{id}=-(1.757 \pm 0.002)$ pF, and 
$V^0_\mathrm{PZT}=(69.31 \pm 0.02)$ V, with a reduced $\chi^2=2.9$.   
The coefficient $A_3^\mathrm{id}$ is in agreement within one standard 
deviation with the less accurate theoretical expectation of 
$-2 \pi \epsilon_0 R=-(1.72 \pm 0.02)$ pF. 
If the $A_2$-terms are not constrained to zero in both fits, one gets 
$A_1^\mathrm{mod}=(223.8 \pm 1.5)$ pF, 
$A_2^\mathrm{mod}=-(359524 \pm 13000)$ pF/m, 
$A_3^\mathrm{mod}=-(433 \pm 135)$ pF, and 
$V^0_\mathrm{PZT}=(68.59 \pm 0.45)$ V, 
with a reduced $\chi^2=13.6$, and   
$A_1^\mathrm{id}=(193.9 \pm 0.2)$ pF, 
$A_2^\mathrm{id}=-(29000 \pm 2800)$ pF/m, 
$A_3^\mathrm{id}=-(1.705 \pm 0.005)$ pF, and 
$V^0_\mathrm{PZT}=(69.25 \pm 0.02)$ V, with a reduced $\chi^2=2.6$.}
\end{figure}

\textit{Distance dependence of the contact potential}: 
In regard to the observation of the dependence on distance of the
contact potential we have reported in \cite{prarckim}, the authors 
of \cite{DeccaComment} show previously unpublished data, in Fig. 3,  
for their experiment located in Indiana. In this plot no systematic trend 
of $V_0$ is observable in the entire explored range of distances.
However an issue arises if the same data, provided to us 
by R. Decca, are plotted including the error bars for $V_0$ at 
one standard deviation, as appearing in Fig. 2.  
The contact potential is not constant within the error bars, showing
scatter over several standard deviations, and perhaps a weak
sinusoidal component. By fitting the data with a constant value,
one gets a reduced $\chi^2$=7.2 (resulting from an absolute
$\chi^2=3,603$ and 499 degrees of freedom for the 500 data
points), therefore the hypothesis of constant contact potential is
highly unlikely, and the use of a constant compensating external
potential will generate large residuals and related systematic
errors. One may argue that the error bars associated to the contact potential have been 
underestimated. In fact, if the error bars are increased by a factor 
$\simeq \sqrt{7.2}$ the reduced $\chi^2$ approaches values of order 
unity, making the constant contact potential hypothesis realistic. 
But doing so increases the average error bar to $\simeq 0.35$ mV, 
with the relative average error in each determination of 
$V_0$ increasing from the initial 0.85 $\%$ to 2.29 $\%$. 
This will invalidate, once propagated through the 
electrostatic calibration analysis, the claimed precision 
of 0.2 $\%$ quoted in \cite{DeccaComment}, in any event 
at variance with the relative error of 0.85 $\%$ resulting from 
the standard deviation of 0.13 mV of $V_0$=15.29 mV quoted in the same paper. 
In other words, this is an example of data already collected 
requiring a reanalysis of the precision and accuracy, as we 
have commented in the conclusions of \cite{prarckim}.
  
It is worth pointing out that the plot presented in Fig. 3 
of \cite{DeccaComment} should not be considered representative 
of the overall picture. In addition to our findings in \cite{prarckim}, 
clear evidence of a distance-dependent $V_0$ has been recently 
found by the groups operating in Grenoble \cite{Jourdanthesis}, Amsterdam
\cite{Iannuzzi}, and Yale \cite{KimLamo}, and in a 
reanalysis of the data collected at the Lucent Laboratories 
\cite{Chan}, as we will report in a future publication. 
Even in the data by the Riverside group reported in Fig. 4 
of \cite{MohideenJPA} a slope seems evident in spite of a 
rather coarse scale chosen for the vertical axis, although 
the authors believe that the scattering of the data 
overwhelms any systematic trend \cite{Mohideenprivate}. 
It is understood that future measurements will clarify under which
conditions the contact potential can be considered as constant
for two surfaces in close proximity to each other, including 
the possible role of image charges \cite{Stipe}, and further investigations on its time 
variability, also started recently \cite{Pollack}, will be necessary.
We also remark on the fact that the presence of a contact potential
depending on the distance requires a proper handling of the
electrostatic term, as recently outlined in \cite{Lamoreauxnote}. 
\begin{figure}[t]
\begin{center}
\includegraphics[width=0.90\columnwidth]{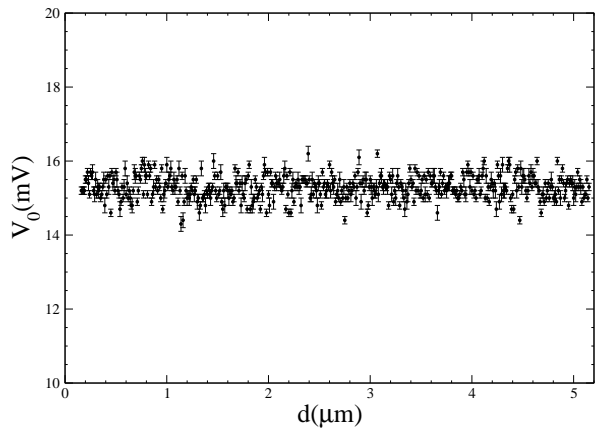}
\end{center}
\caption{Minimizing voltage versus sphere-cantilever distance resulting
for the data analysis of the electrostatic calibration for the same 
run as in Fig. 3 of \cite{DeccaComment}, also including the error 
bars of each determination of $V_0$ (courtesy of R. Decca). 
By considering the scattering of the data around their average value
of $\langle V_0 \rangle$=15.29 mV we obtain a standard deviation of 
$\Delta V_0$= 0.31 mV (at variance with the quoted value of 0.13 mV in 
Fig. 3 of \cite{DeccaComment}), while the average value of the error 
bars is $\delta V_0$=0.13 mV. See the text for the discussion of the 
$\chi^2$ analysis under the hypothesis of a constant value for the 
contact potential.} 
\end{figure}

\textit{General issues on precision and accuracy}:  The authors of
\cite{DeccaComment} also comment about the claimed precision and 
accuracy of the experiments in the sphere-plane configuration, and 
on the negligible role of the patch potentials in their experimental setups.
We believe that various claims on the precision and accuracy in 
previous papers - and the mixing of experimental facts and theoretical 
hypotheses supposed to be tested by the data themselves - have generated 
confusion in the literature, and inconsistent methodologies for their 
quantitative assessments. As a first representative example, we mention 
\cite{Mohideen} (see page 4552, first column) in which the precision of the measurement has 
been assessed by comparing the data with the expected Casimir force.  
This generates a manifestly model-dependent experimental precision 
in contradiction with the concept that this quantity is merely 
a figure of merit of the reproducibility of the measurements. 
Also, the model for patch potentials in \cite{Speake} assumes some specific
form for the two-dimensional Fourier spectrum of patches and crucially
depends on the range of spatial wavelengths contributing to it. Using
this phenomenological model and a spectral range arbitrarily chosen 
based on the size of the grains in the material, in \cite{DeccaComment} 
it is concluded that the effects of patches have been investigated in
detail in \cite{DeccaAnn} and \cite{Mohideen2004} and found negligible
in those experiments. On the contrary, we believe that the model in 
\cite{Speake} does not necessarily describe any realistic experiment
(as indeed argued in \cite{Speake} itself) and that a {\it detailed} 
investigation of patch effects in Casimir experiments would require 
in-situ measurements, using for instance Kelvin probe techniques \cite{Robertson}.   
A third example of model-dependent experimental claim in the same 
spirit is the evaluation of the limits to Yukawian forces based on 
a formal mapping of the sphere-plane configuration to the parallel
plane case via PFA \cite{DeccaAnn} (see page 74) already mentioned in the introduction.
All these methodological issues will also require further in depth 
analysis, both on the data collected so far, and via a third generation 
of experiments capitalizing on the current discussions. 

In conclusion, we believe that the message in our contribution
\cite{prarckim} is reinforced by the discussion presented in
\cite{DeccaComment}: a very detailed geometrical and electrical
characterization of the sphere-plane setup is necessary prior to
assessing the precision and accuracy of the Casimir force
measurements. Characterizations of this nature require dedicated 
efforts, as for instance initiated in \cite{Palas} where systematic 
measurements have been carried out to study the dependence of the 
forces upon the shape and roughness of the microspheres, or 
using apparata with unprecedented levels of stability \cite{Iannuzzi}.
It is also crucial to declare all the relevant parameters 
in the entire explored range of distances as first stressed in \cite{IannuzziPNAS}, 
including their possible time variability, and also quoting the
percentage of rejection of samples or data runs. 

\newpage

\noindent
$^*${Present address: Department of Physics, Yale University,
  217 Prospect Street, New Haven, CT 06520-8120.}

\vspace{0.1in}

\noindent
DARD and RO are grateful to R. Decca for providing data of his experiment, 
and to R. Decca and U. Mohideen for stimulating discussions.
WJK is grateful to T. Emig and R. Podgornik for useful discussions.

\end{document}